\documentclass[namedreferences]{solarphysics}

\usepackage[hyperref,optionalrh,showbiblabels]{spr-sola-addons} 
\usepackage{graphicx}        
\usepackage{color}           
\usepackage{breakurl}        
\usepackage{lineno}
\usepackage[T1]{fontenc}
\ifx\urlurl    \undefined
  \def\urlurl#1{\href{http://#1}{\textsf{#1}}}
\fi

\chardef\us=`\_

\begin{document}
\begin{article}
\begin{opening}

\title{Spectral and Imaging Observations of a C2.3 White-Light Flare from the \emph{Advanced Space-Based Solar Observatory} (ASO-S) and the \emph{Chinese H$\alpha$ Solar Explorer} (CHASE)}

\author[addressref={aff1,aff2}]{\inits{Q.}\fnm{Qiao}~\lnm{Li} \orcid{0000-0001-7540-9335}}
\author[addressref={aff1,aff2},corref,email={yingli@pmo.ac.cn}]{\inits{Y.}\fnm{Ying}~\lnm{Li}\orcid{0000-0002-8258-4892}}
\author[addressref={aff1,aff2}]{\inits{Y.}\fnm{Yang}~\lnm{Su}\orcid{0000-0002-4241-9921}}
\author[addressref={aff1}]{\inits{D.}\fnm{Dechao}~\lnm{Song}\orcid{0000-0003-0057-6766}}
\author[addressref={aff1,aff2}]{\fnm{Hui}~\lnm{Li}\orcid{0000-0003-1078-3021}}
\author[addressref={aff1,aff2}]{\fnm{Li}~\lnm{Feng}\orcid{0000-0003-4655-6939}}
\author[addressref={aff1,aff2}]{\fnm{Yu}~\lnm{Huang}\orcid{0000-0002-0937-7221}}
\author[addressref={aff1,aff2}]{\fnm{Youping}~\lnm{Li}\orcid{0000-0001-5529-3769}}
\author[addressref={aff1}]{\fnm{Jingwei}~\lnm{Li}}
\author[addressref={aff1}]{\fnm{Jie}~\lnm{Zhao}\orcid{0000-0003-3160-4379}}
\author[addressref={aff1}]{\fnm{Lei}~\lnm{Lu}\orcid{0000-0002-3032-6066}}
\author[addressref={aff1}]{\fnm{Beili}~\lnm{Ying}\orcid{0000-0001-8402-9748}}
\author[addressref={aff1}]{\fnm{Jianchao}~\lnm{Xue}\orcid{0000-0003-4829-9067}}
\author[addressref={aff1}]{\fnm{Ping}~\lnm{Zhang}}
\author[addressref={aff1,aff2}]{\fnm{Jun}~\lnm{Tian}\orcid{0000-0002-1068-4835}}
\author[addressref={aff1,aff2}]{\fnm{Xiaofeng}~\lnm{Liu}\orcid{0000-0002-3657-3172}}
\author[addressref={aff1,aff2}]{\fnm{Gen}~\lnm{Li}}
\author[addressref={aff1,aff2}]{\fnm{Zhichen}~\lnm{Jing}\orcid{0000-0002-8401-9301}}
\author[addressref={aff1,aff2}]{\fnm{Shuting}~\lnm{Li}}
\author[addressref={aff1,aff2}]{\fnm{Guanglu}~\lnm{Shi}\orcid{0000-0001-7397-455X}}
\author[addressref={aff1,aff2}]{\fnm{Zhengyuan}~\lnm{Tian}\orcid{0000-0002-2158-0249}}
\author[addressref={aff1,aff2}]{\fnm{Wei}~\lnm{Chen}\orcid{0000-0002-4118-9925}}
\author[addressref={aff1,aff2}]{\fnm{Yingna}~\lnm{Su}\orcid{0000-0001-9647-2149}}
\author[addressref={aff1,aff2}]{\fnm{Qingmin}~\lnm{Zhang}\orcid{0000-0003-4078-2265}}
\author[addressref={aff1,aff2}]{\fnm{Dong}~\lnm{Li}\orcid{0000-0002-4538-9350}}
\author[addressref={aff1}]{\fnm{Yunyi}~\lnm{Ge}}
\author[addressref={aff1,aff2}]{\fnm{Jiahui}~\lnm{Shan}}
\author[addressref={aff1,aff2}]{\fnm{Yue}~\lnm{Zhou}\orcid{0000-0002-3341-0845}}
\author[addressref={aff1}]{\fnm{Shijun}~\lnm{Lei}}
\author[addressref={aff1,aff3}]{\fnm{Weiqun}~\lnm{Gan}\orcid{0000-0001-9979-4178}}

\address[id=aff1]{Key Laboratory of Dark Matter and Space Astronomy, Purple Mountain Observatory, Chinese Academy of Sciences, Nanjing 210023, People’s Republic of China}
\address[id=aff2]{School of Astronomy and Space Science, University of Science and Technology of China, Hefei 230026, People’s Republic of China}
\address[id=aff3]{University of Chinese Academy of Sciences, Nanjing 211135, China}

\runningauthor{Q. Li et al.}
\runningtitle{ASO-S and CHASE Observations of a C2.3 WLF}

\begin{abstract}
Solar white-light flares are characterized by an enhancement in the optical continuum, which are usually large flares (say X- and M-class flares). Here we report a small C2.3 white-light flare (SOL2022-12-20T04:10) observed by the \emph{Advanced Space-based Solar Observatory} and the \emph{Chinese H$\alpha$ Solar Explorer}. This flare exhibits an increase of $\approx$6.4\% in the photospheric Fe \textsc{i} line at 6569.2\,\AA\ and {$\approx$3.2\%} in the nearby continuum. The continuum at 3600\,\AA\ also shows an enhancement of $\approx$4.7\%. The white-light brightening kernels are mainly located at the flare ribbons and co-spatial with nonthermal hard X-ray sources, which implies that the enhanced white-light emissions are related to nonthermal electron-beam heating. At the brightening kernels, the Fe \textsc{i} line displays an absorption profile that has a good Gaussian shape, with a redshift up to $\approx$1.7 km s$^{-1}$, while the H$\alpha$ line shows an emission profile though having a central reversal. The H$\alpha$ line profile also shows a red or blue asymmetry caused by plasma flows with a velocity of several to tens of km s$^{-1}$. It is interesting to find that the H$\alpha$ asymmetry is opposite at the conjugate footpoints. It is also found that the CHASE continuum increase seems to be related to the change of photospheric magnetic field. Our study provides comprehensive characteristics of a small white-light flare that help understand the energy release process of white-light flares.
\end{abstract}

\keywords{Flares, White-Light, Relation to Magnetic Field, X-Ray Bursts, Spectral Line, Continuum}
\end{opening}


\section{Introduction}
     \label{S-Introduction} 

Solar white-light flares (WLFs) are identified by a sudden increase in the visible continuum \citep{Neidig&Cliver1983, Neidig1989}. The first WLF (also the first solar flare) was reported by \citet{Carrington1859} and \citet{Hodgson1859}. Most WLFs occur in the vicinity of sunspots and manifest as brightening kernels, but some can also occur in a region almost without sunspot \citep[e.g.][]{Hudson1994} and appear as loop-like structures (e.g. \citealt{Hudson1992, Hudson2006}; \citealt{Jejcic2018}). While it is generally accepted that WLFs constitute a minority in the overall flare family \citep{Fang2013} with X- and M-class flares (i.e. major flares) being the most energetic, there are different views suggesting that all flares may exhibit white-light brightenings \citep[e.g.][]{Matthews2003, Jess2008, Song2018aa}. This hypothesis gains support from the discovery of white-light continuum enhancements in small flares.  

Up to now, there have been only about 20 C-class WLFs reported including several low C-class ones from both statistical studies \citep[e.g.][]{Matthews2003, Hudson2006, Song2018aa, Song2018ApJ, Castellanos_Duran2020} and case analyses \citep[e.g.][]{Jess2008, Song2020}. Statistically, C-class WLFs exhibit an average increase of $\approx$10\% in the visible continuum. Some C-class WLFs with a lower magnitude could display a relatively higher white-light enhancement \citep[e.g.][]{Hudson2006, Song2018aa, Song2018ApJ}, which seems to deviate from the behavior as observed in major flares. Note that this discrepancy may be attributed to sample bias and instrumental limitation. For case studies, \citet{Jess2008} observed a C2.0 WLF showing a white-light kernel with a diameter of $\approx$0.4$^{\prime\prime}$ using the \emph{Swedish Solar Telescope} (SST) among the highest resolution telescopes. This small scale WLF demonstrates an unusually high increase of $\approx$300\% in the blue continuum, surpassing a typical enhancement in WLFs. However, degrading the spatial resolution to 1$^{\prime\prime}$ decreases the enhancement to only 1\%. In another case study, \cite{Song2020} presented a C2.3 WLF with an enhancement of $\approx$18\% observed by the \emph{Helioseismic and Magnetic Imager} \citep[HMI:][]{Scherrer2012} on the \emph{Solar Dynamics Observatory} \citep[SDO:][]{Pesnell2012}. This particular flare is associated with weak hard X-ray (HXR) emissions, with the white-light enhancement unlikely caused by accelerated electron beams. Its magnetic field morphology and evolution are more supportive of magnetic reconnection in the lower atmosphere \citep[e.g.][]{Ding1999, Chen2001} rather than in the corona. This suggests that both the non-thermal particle injection and evolution of magnetic fields need to be considered in studying WLFs. Given the scarcity of studies on small WLFs, more attention should be paid to the WLFs with a low magnitude. 

Multiple studies have shown a good correlation between the white-light (including the Balmer continuum) and HXR emissions, in terms of both time and space (e.g. \citealt{Matthews2003, Metcalf2003, Hudson2006, Wang2009, Krucker2011, MartinezOliveros2012}; \citealt{Heinzel2014, Heinzel2017}). These results support an electron-beam bombardment \citep[e.g.][]{Hudson1972, Aboudarham1987} plus its subsequent chromospheric condensation \citep[e.g.][]{Gan1992, Gan1994, Kowalski2015} and radiative backwarming \citep[e.g.][]{Machado1989} processes being related to the white-light brightenings. Some studies further reveal that the footpoint or ribbon with a stronger white-light enhancement shows a stronger HXR flux \citep[e.g.][]{Metcalf2003, Krucker2011}. However, \citet{{Chen2005}} reported a stronger white-light kernel accompanied by a weaker nonthermal energy flux in comparison between two footpoints. Even some cases show no relationship between the white-light and HXR emissions \citep[e.g.][]{Ryan1983, Ding1994, Sylwester2000, Song2020}. On the other hand, some other heating mechanisms can also play a role in WLFs, such as magnetic reconnection in the lower atmosphere \citep[e.g.][]{Ding1999, Chen2001}, energy dissipation of Alfv\'{e}n waves \citep[e.g.][]{Emslie1982, Fletcher2008}, and soft X-ray (SXR)/ultraviolet (UV) irradiation \citep[e.g.][]{Poland1988}. 

Previous studies have indicated that permanent changes in the photospheric magnetic field commonly appear in flares \citep[e.g.][]{Sudol2005, Castellanos_Duran2018}. These changes mean that the magnetic components undergo abrupt transformations and do not return to their original levels for a long time. The coronal implosion process \citep{Hudson2000} predicts that photospheric fields become more horizontal during flares, thereby changing the line-of-sight (LoS) magnetic field strength [$B_{\rm LoS}$] \citep{Hudson2008}. This has been used for an interpretation of the appearance of $B_{\rm LoS}$ changes [$\Delta B_{\rm LoS}$] \citep[e.g.][]{Petrie2010, Gosain2012, Sun2012}. However, differences in $\Delta B_{\rm LoS}$ between the chromosphere and photosphere suggested that the chromospheric $\Delta B_{\rm LoS}$ may not support a contraction of the field line in flare loops \citep[e.g.][]{Kleint2017}. Recently, a statistical study \citep{Castellanos_Duran2020} showed that white-light brightenings and $\Delta B_{\rm LoS}$ in the photosphere can overlap less than 60\%, and the areas of these two parameters appear to have a power-law relation. The relationship between $\Delta B_{\rm LoS}$ and white-light brightenings deserves a further study, especially for small WLFs.

Spectral analysis is a good way to study the dynamics of plasma in WLFs, especially via some Fe \textsc{i} lines formed in the photosphere \citep[e.g.][]{Babin1992, Lin1996, Hong2018}. \citet{Jurcak2018} reported a WLF with the response of Fe \textsc{i} lines at 6301.5\,\AA\ and 6302.5 \AA, which was estimated to originate in the chromosphere instead of photosphere. In addition, the H$\alpha$ line at 6562.8\,\AA\ formed from the photosphere to chromosphere can help us understand the physical process of WLFs. This line mainly behaves as an emission profile with a central reversal and a red asymmetry during WLFs \citep[e.g.][]{Babin1992, Zhou1997}.

In this work, we report a small C2.3 WLF (SOL2022-12-20T04:10) observed by the \emph{Advanced Space-based Solar Observatory} \citep[ASO-S:][]{Gan2023} and the \emph{Chinese H$\alpha$ Solar Explorer} \citep[CHASE:][]{Li2022}. ASO-S provides images in the white-light continuum at 3600\,\AA\ and also HXR spectra at $\approx$10\,--\,300 keV. CHASE has spectral observations in the Fe \textsc{i} 6569.2\,\AA\ and H$\alpha$ 6562.8\,\AA\ lines. All these allow us to study the comprehensive characteristics of this C2.3 WLF when combining with some other multi-wavelength observations. In the following Section \ref{S-Data}, we describe the data reduction. Our results are presented in Section \ref{S-Results}, followed by a summary and discussion in Section \ref{S-Summary}.

\section{Data Reduction}
     \label{S-Data} 

\subsection{Data}

The data used in this work are obtained by multiple telescopes. The main telescopes include the \emph{Lyman-alpha Solar Telescope} \citep[LST:][]{Li2019, Chen2019, Feng2019} aboard ASO-S, CHASE, and HMI on SDO. 
LST consists of three instruments, one of which is the \emph{White-light Solar Telescope} (WST). WST can provide full-disk images in the continuum at 3600$\pm$20 \AA. The pixel size of the images is $\approx$0.5$^{\prime\prime}$ while the spatial resolution is about 4$^{\prime\prime}$ \citep{Jing2024}. The cadence is two minutes in a routine mode. The level-2.5 (L2.5) data of WST are used here, which have been done for scientific calibration and image registration. CHASE provides spectral data of the Fe \textsc{i} line at 6569.2\,\AA\ (6567.8\,--\,6570.6 \AA) and the H$\alpha$ line at 6562.8\,\AA\ (6559.7\,--\,6565.9 \AA) via a raster scanning mode (RSM). The data have a cadence of $\approx$1 minute, a pixel size of $\approx$1$^{\prime\prime}$, and a spectral resolution of 0.048 \AA\ pixel$^{-1}$, with a two-binning mode in both space and spectrum. The calibrated L1.5 data are used here \citep{Qiu2022}. The full-disk images of Fe \textsc{i} 6173\,\AA\ and the LoS magnetogram presented in the work are obtained from SDO/HMI, which have a pixel size of 0.5$^{\prime\prime}$ and a cadence of 45\,seconds.

The \emph{Hard X-ray Imager} \citep[HXI:][]{Zhang2019, Su2022} aboard ASO-S is an HXR imaging spectrometer observing the Sun in an energy range of $\approx$10\,--\,300 keV with energy and angular resolutions of $\approx$16.5\% at 32 keV and 3.1$^{\prime\prime}$, respectively. The HXI L1 data are used in the present study. The \emph{Spectrometer Telescope for Imaging X-rays} \citep[STIX:][]{Krucker2020} on \emph{Solar Orbiter} \citep[SolO:][]{Muller2020} has energy and angular resolutions of $\approx$1 keV @6 keV and $\geq$7$^{\prime\prime}$, respectively, in a range from 4 keV to 150 keV. Its L1 pixel and spectrogram data are used here. At the observation time of the flare event, STIX had a separation angle of 19.2$^{\circ}$ with Earth and a distance of 0.9 AU to the Sun. The \emph{Atmospheric Imaging Assembly} \citep[AIA:][]{Lemen2012} on SDO provides the images for flare loops in the Extreme ultraviolet (EUV) channel at 131 \AA\, with a pixel size of 0.6$^{\prime\prime}$ and a cadence of 12 seconds. The SXR 1\,--\,8 Å light curve is from the \emph{X-Ray Sensor} \citep[XRS:][]{Hanser1996} aboard the \emph{Geostationary Operational Environmental Satellite} (GOES) with a high cadence of 1 second.

\subsection{Data Reduction} 
\label{S-redu}

We make a co-alignment for the images from different instruments via the routines in the SolarSoftWare (SSW: \citealt{Freeland1998}). Firstly, the AIA and HMI images are registered using \textsf{aia\_prep.pro}. The solar rotation has been removed via \textsf{drot\_map.pro}. Then the white-light images from WST and CHASE are co-aligned with the HMI continuum images using \textsf{xyoff.pro} based on cross correlation for sunspot features. The uncertainty of the co-alignment is estimated to be about 1$^{\prime\prime}$.

For HXI light curves at 20\,--\,50 keV and 50\,--\,100 keV, we use the combined data of three total flux detectors (D92, D93, and D94) with a cadence of four seconds. The images are reconstructed using HXI CLEAN algorithm \citep{Su2019} with sub-collimators G3\,--\,G10, which correspond to a spatial resolution of $\approx$6.5$^{\prime\prime}$. We also use STIX data to produce the light curves at 4\,--\,10 keV and 10\,--\,20 keV with a cadence of four seconds and to reconstruct images via the Expectation Maximization (EM) algorithm \citep{Massa2019}. In addition, we make spectral fitting from STIX via a thermal component plus a nonthermal thick target component. Note that the light-travel time at Earth with a correction of 35 s for STIX has been considered.

In order to remove non-flaring contributions and calculate the relative enhancements caused by flare, we subtract a background or base image from flaring images for different wave bands. For HMI and WST images, an average for 30 minutes before the flare onset is adopted as the background. For CHASE images, the last moment at 04:19:48 UT, i.e. around the flare end time, is selected as the background which is relatively quiet during the whole observation of CHASE. Note that this may lead to an underestimation of the CHASE continuum enhancement to some extent. When estimating the uncertainty of relative enhancements for each wave-band emission, we choose a quiet-Sun region with a size of 100$^{\prime\prime}$$\times$16$^{\prime\prime}$ ($X=[-799^{\prime\prime}, -699^{\prime\prime}]$, $Y=[483^{\prime\prime}, 499^{\prime\prime}]$) to calculate its relative intensity fluctuations, i.e. the error bars as plotted in Figures \ref{fig:3}a and b. Note that this quiet-Sun region is out of the field of view of the images as shown in Figures \ref{fig:1}d\,--\,i.

Using the CHASE spectral data, we can measure the Doppler velocities of Fe \textsc{i} and H$\alpha$ lines and also calculate the H$\alpha$ asymmetry. The reference line centers of Fe \textsc{i} and H$\alpha$ are determined by averaging the profiles over the quiet-Sun region as mentioned above. The excess profile, i.e. after subtracting the background profile, is adopted to obtain the Doppler velocity. For Fe \textsc{i}, a single Gaussian fit with a linear background is used and the average of the background is further used to define the nearby continuum. It is found that the intensity of this defined continuum is approximately equal to that at the wavelength of 6568.6 \AA. For H$\alpha$, bisector \citep[e.g.][]{Chae2013} and moment analyses are applied to derive its Doppler velocity. The velocity uncertainties of Fe \textsc{i} and H$\alpha$ lines are estimated to be within $\pm$0.6 and $\pm$3.9 km s$^{-1}$, respectively. The H$\alpha$ asymmetry is calculated based on its original profiles. Referring to Equation (1) in \citet{Asai2012}, the asymmetry is expressed as follows, 
\begin{equation}  \label{Eq-1}
RA = \frac{I_{\rm rp}-I_{\rm bp}}{{I_{\rm rp}+I_{\rm bp}}}, 
\end{equation}
where $I_{\rm rp}$ and $I_{\rm bp}$ denote the peaks of H$\alpha$ red and blue wings, respectively. A positive/negative {\em RA} represents a red/blue asymmetry. The uncertainty of the asymmetry is estimated to be about $\pm$0.014. When showing the profiles of Fe \textsc{i} and H$\alpha$ (in Figures \ref{fig:3}c\,--\,f and \ref{fig:4}a\,--\,d), we make a normalization by using the nearby quiet-Sun continuum, i.e. in a contrast way. More specifically, the nearby quiet-Sun continuum intensity of the H$\alpha$ line is determined by averaging the intensities of the ten wavelength points in its bluest wing, while that of the Fe \textsc{i} line is defined as the average of the linear background of this line.

To investigate the relationships of white-light emission with $B_{\rm LoS}$ and $\Delta B_{\rm LoS}$ in the photosphere, the magnetograms from HMI are reshaped into the same pixel size as CHASE images. Then a stepwise function \citep{Sudol2005} describes the temporal evolution of $B_{\rm LoS}$ as expressed by
\begin{equation} \label{Eq-2}
B_{\rm LoS}(t) = a + bt + c\left\{ 1 + \frac{2}{\pi}tan^{-1}[n(t-t_{0})]\right\},
\end{equation}
in which $t$\ represents time and $a$, $b$, $c$, $n$, and $t_0$ are free parameters to fit. From this stepwise function, the characterized start time [${t_s = t_0 - \pi n^{-1}}$] and end time [${t_e = t_0 + \pi n^{-1}}$] can also be obtained. $\Delta B_{\rm LoS}$ is derived mainly via the method in \cite{Castellanos_Duran2018}, which could be expressed as:
\begin{equation} \label{Eq-3}
    \Delta B_{\rm LoS}=\left\{
	\begin{array}{lc}
		sgn(t_{\rm max}-t_{\rm min})[B_{\rm LoS}(t_{\rm max})-B_{\rm LoS}(t_{\rm min})]\\
		2c\\
            B_{\rm LoS}(t_e)-B_{\rm LoS}(t_s)-2\pi bn^{-1}
	\end{array}
	\right., 
\end{equation}
where $t_{\rm max}$ and $t_{\rm min}$ represent the times of the maximum and minimum of $B_{\rm LoS}$, respectively, and the function $sgn(x)$ represents the positive or negative sign of $x$. Here we also modify some criteria as follows. 
\begin{enumerate}
\item{In this short-duration event, the time range of fit is reduced to 60 minutes centered at the GOES peak time.}
\item{The duration parameter [$\pi n^{-1}$] must be smaller than the fit period, otherwise $\Delta B_{\rm LoS}$ = 0.}
\item{$t_s$ must be earlier than $t_e$, otherwise $\Delta B_{\rm LoS}$ = 0.}
\item{Both $t_s$ and $t_e$ must be within the fit period, otherwise $\Delta B_{\rm LoS}$ = 0.}
\end{enumerate}
Note that the noise level of $B_{\rm LoS}$ and $\Delta B_{\rm LoS}$ is determined as about ten Gauss \citep{Liu2012}. 

\section{Results}
     \label{S-Results} 

\subsection{Overview of the C2.3 Event}
  \label{S-lc}
  
\begin{figure} 
\centerline{\includegraphics[width=1 \textwidth]{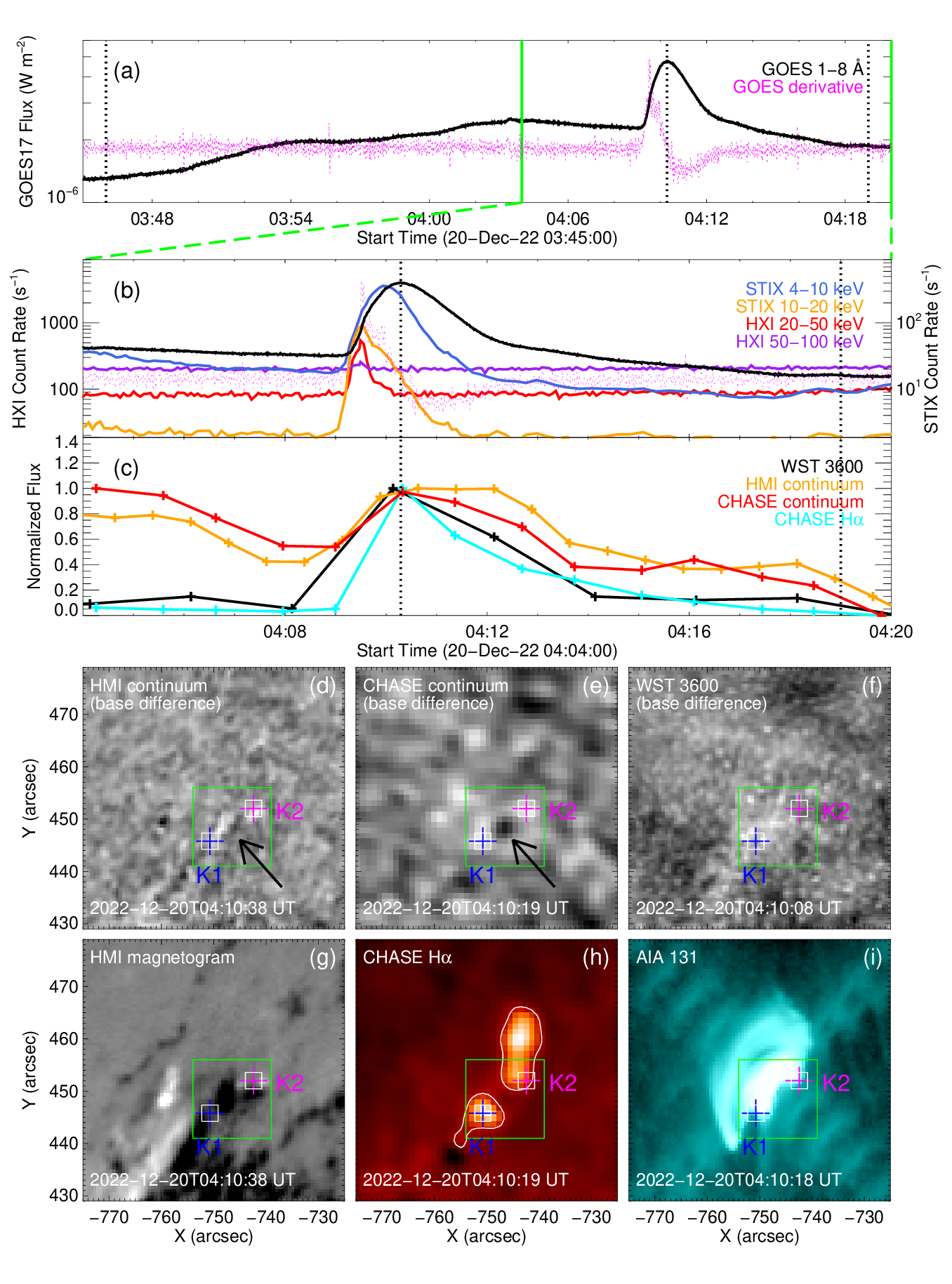}}
\caption{(\textbf{a}) Full-disk GOES SXR 1\,--\,8\,\AA\ light curve and its temporal derivative for the C2.3 WLF. The \emph{three vertical dotted lines} mark the GOES onset, peak, and end times. The \emph{two green vertical lines} denote the CHASE observation period as shown in Panels \textbf{b} and \textbf{c}. (\textbf{b}) Full-disk light curves of STIX HXR 4\,--\,10 keV, STIX HXR 10\,--\,20 keV, HXI HXR 20\,--\,50 keV, and HXI HXR 50\,--\,100 keV, together with GOES SXR 1\,--\,8\,\AA\ flux and its temporal derivative. (\textbf{c}) Normalized light curves in multiple wavebands from WST, HMI, and CHASE continua and CHASE H$\alpha$ line (summed over $\pm$2.3 \AA). The \emph{two vertical dotted lines} in Panels \textbf{b} and \textbf{c} represent the GOES peak and end times. (\textbf{d\,--\,i}) Base-difference images of HMI, CHASE, and WST continua, HMI magnetogram, CHASE H$\alpha$ image, and AIA 131 image near the flare peak time. The \emph{two pluses} in each panel denote the two white-light brightening kernels K1 and K2. The \emph{green box} in Panels \textbf{d\,--\,i} marks the flaring region used to make the light curves in Panel \textbf{c}. The \emph{black arrow} in Panels \textbf{d} and \textbf{e} denotes a small pore. The \emph{white contours} in Panel \textbf{h} outline the H$\alpha$ flare ribbons whose excess intensities are larger than 500 DNs.}\label{fig:1}
\end{figure}

The event under study is a GOES C2.3 flare (SOL2022-12-20T04:10) occurring in NOAA active region 13171 (N26E59). This flare began at 03:46 UT, peaked at 04:10 UT, and ended at 04:19 UT on 20 December 2022 (see the GOES SXR 1\,--\,8 \AA\ light curve as shown in Figure \ref{fig:1}a), showing a relatively long slow rise before 04:04 UT. CHASE observations covered the main period of the flare from 04:04 to 04:20 UT (denoted by two green vertical lines in Figure \ref{fig:1}a), which we focus on in the following. Figure \ref{fig:1}b plots the HXR light curves at different energy bands from STIX and HXI, together with the SXR 1\,--\,8 \AA\ flux and its temporal derivative. We can see that the HXR 10\,--\,20 as well as 20\,--\,50 keV emissions peak around the same time as the SXR temporal derivative, indicating the Neupert effect \citep{Neupert1968}. Note that the flare has no response in HXR emission above $\approx$50 keV. Figure \ref{fig:1}c shows multiple emission curves integrated over the flaring core region as marked by the green box in Figures \ref{fig:1}d\,--\,i. It is seen that the emissions at WST 3600 \AA\ and CHASE H$\alpha$ exhibit a notable rise followed by a decay during the flare. Note that their emission peaks may not represent the true peaks due to a low cadence of more than one minute. It is interesting that the CHASE and HMI continuum emissions show a similar trend, though having multiple peaks during the observation period. It should be mentioned that the main peaks of CHASE and HMI emissions appear in the main phase of the flare. 

Figures \ref{fig:1}d\,--\,i display multi-wavelength images of the flaring region around the flare peak time. We find two main white-light brightening kernels, referred to as K1 and K2, from the base-difference images of HMI, CHASE, and WST continua (Figures \ref{fig:1}d\,--\,f). These two kernels are located near a small pore (indicated by a black arrow) as seen in the HMI and CHASE continuum images in Figures \ref{fig:1}d and e. From the HMI magnetogram in Figure \ref{fig:1}g, we can see that K1 is located in negative polarities with a relatively strong magnetic field, while K2 seems to lie in a mix of positive and negative polarities with a weak field, which may be influenced by a projection effect. These two kernels are co-spatial with the H$\alpha$ ribbons as shown in Figure \ref{fig:1}h. K1 is located in the center of the south ribbon with a bright H$\alpha$ source, while K2 lies on the edge of the north ribbon with a relatively weak H$\alpha$ brightening. From the AIA 131\,\AA\ image in Figure \ref{fig:1}i, one can see that K1 and K2 connect the same set of flare loops, which are likely to be a pair of conjugate footpoints. 

\subsection{Spatial Relationship of the White-light Kernels with Nonthermal HXR Sources} 
  \label{S-hxr}

\begin{figure} 
\centerline{\includegraphics[width=1 \textwidth]{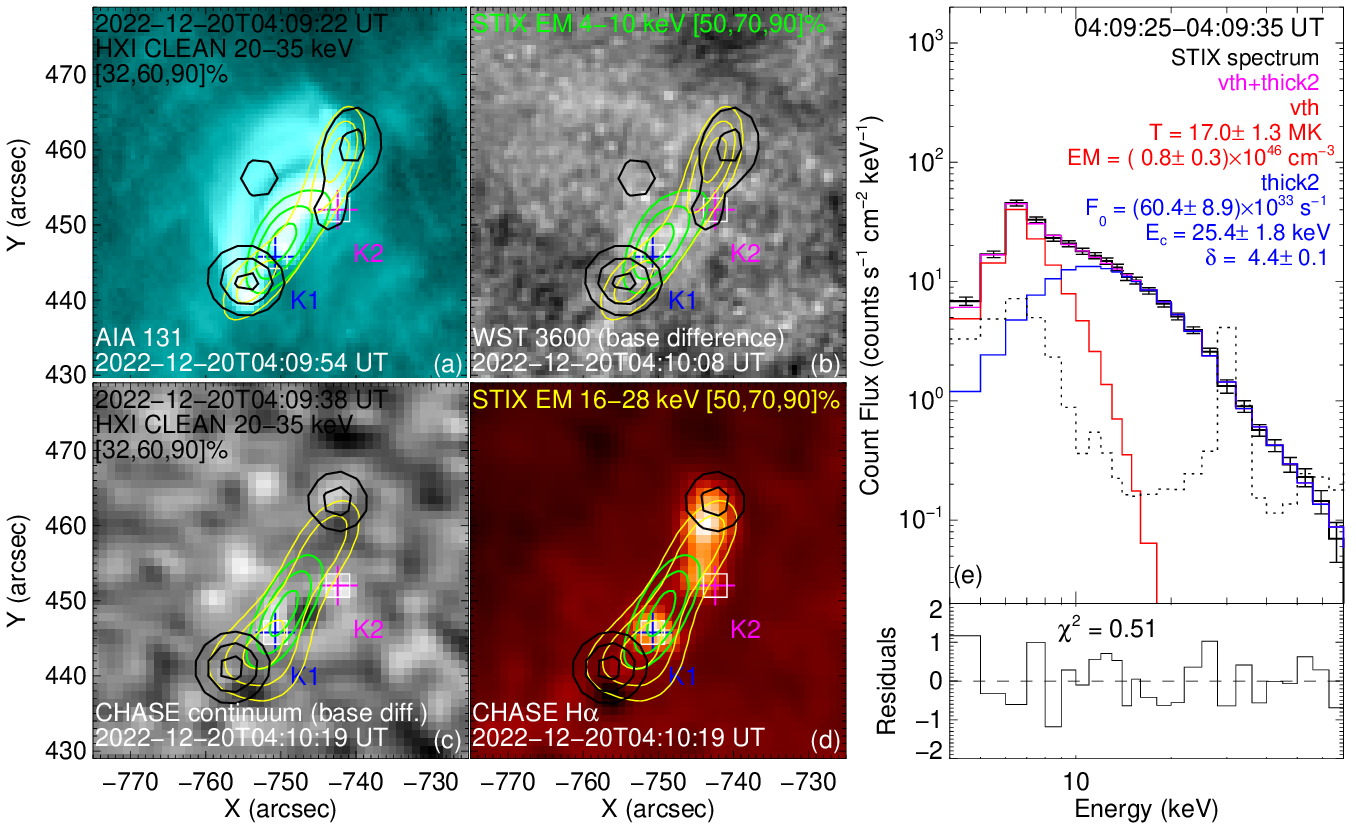}}
\caption{(\textbf{a\,--\,d}) HXR imaging from HXI and STIX for two times overlaid on AIA 131 \AA, WST 3600 \AA, CHASE continuum, and CHASE H$\alpha$ images. The nonthermal and thermal sources are marked by the \emph{black} (HXI 20\,--\,35 keV), \emph{yellow} (STIX 16\,--\,28 keV), and \emph{green} (STIX 4\,--\,10 keV) contours. The \emph{two pluses} in each panel represent the two white-light brightening kernels K1 and K2. (\textbf{e}) STIX HXR background-substracted count spectrum (\emph{black solid curve}) fitted by an isothermal (\emph{red curve}) plus a nonthermal thick target (\emph{blue curve}) model for the peak time of the HXR 20\,--\,50 keV emission. The \emph{gray dashed curve} is the background emission taken during non-flaring times close to the event.}\label{fig:2}
\end{figure}

Figure \ref{fig:2} shows the HXR spectral imaging and fitting from STIX and HXI for the C2.3 flare. From the spectral fitting (Figure \ref{fig:2}e), we can see a nonthermal component above $\approx$20 keV. Thus, we reconstruct the HXR images at 20\,--\,35 keV from HXI for the main nonthermal sources. In addition, we reconstruct the HXR images at 16\,--\,28 keV and 4\,--\,10 keV from STIX for nonthermal and also thermal sources. These HXR sources corresponding to the flare ribbons and also loops are overplotted on AIA 131 \AA, WST 3600 \AA, and CHASE continuum and H$\alpha$ images (Figures \ref{fig:2}a\,--\,d). At an earlier time of 04:09:22 UT (Figures \ref{fig:2}a and b), the nonthermal HXR sources at 20\,--\,35 keV are mainly cospatial with the flare ribbons. In particular, the two white-light kernels, K1 and K2, match the nonthermal sources well. It is noted that K1 is close to the nonthermal source with a larger HXR flux while K2 corresponds to the one with a lower flux. At a later time of 04:09:38 UT (Figures \ref{fig:2}c and d), the HXR sources evolve a little while K1 is still close to one of the nonthermal HXR sources. These results suggest that the white-light emissions are related to a nonthermal electron-beam heating although the timings of white-light emissions and HXR sources are not conclusive due to a low cadence of the white-light data from WST and CHASE.

\subsection{Spectral Features of the White-light Kernels} 
  \label{S-spec} 
  
\begin{figure} 
\centerline{\includegraphics[width=1 \textwidth]{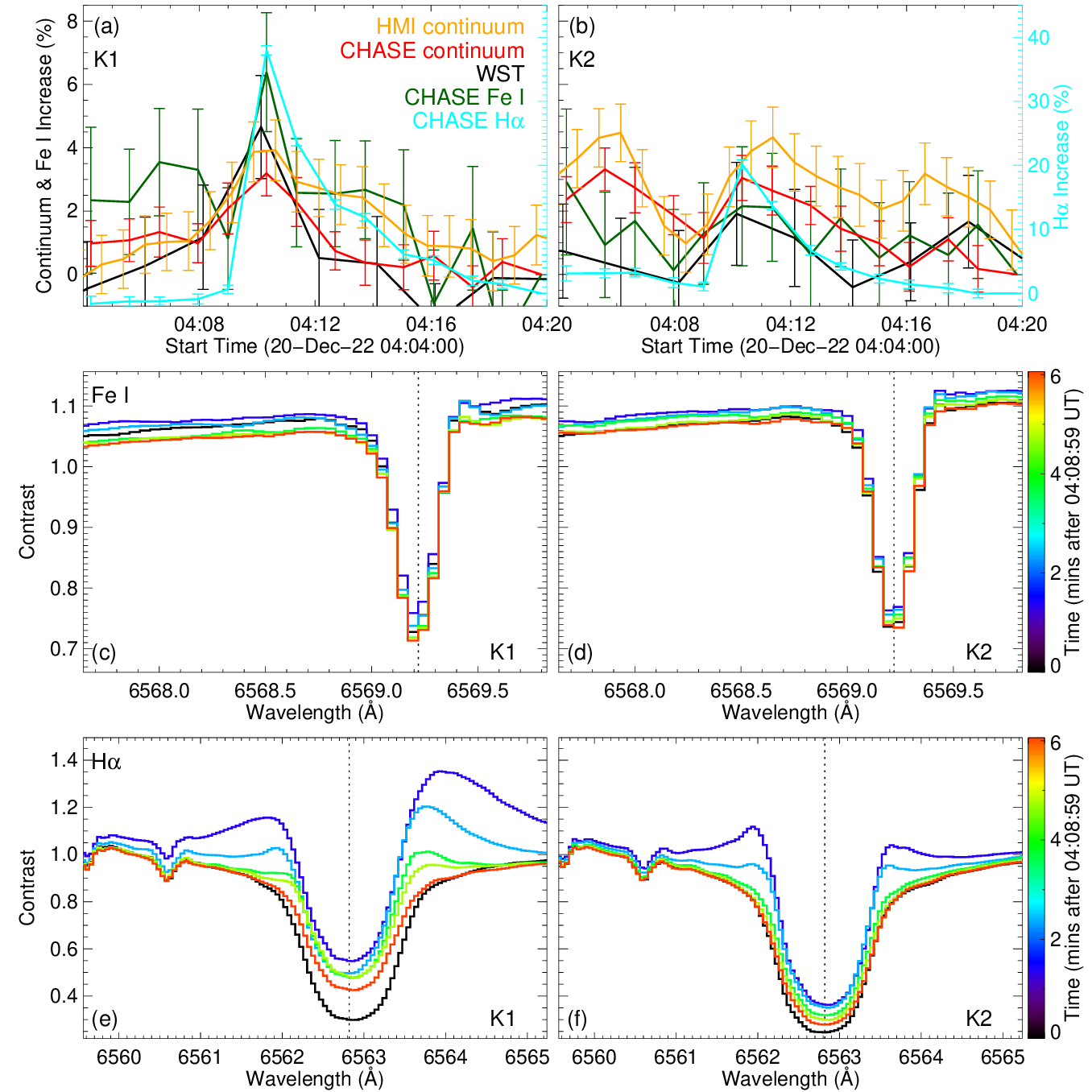}}
\caption{(\textbf{a}) and (\textbf{b}) Temporal evolution of the relative enhancements or increases of HMI continuum, CHASE continuum, WST continuum, Fe \textsc{i} line core, integrated H$\alpha$ line ($\pm$2.3 \AA) at the white-light kernels K1 and K2 averaged over an area of 3$^{\prime\prime}$$\times$3$^{\prime\prime}$. The \emph{error bars} represent the uncertainties. (\textbf{c\,--\,f}) Temporal evolution of the normalized Fe \textsc{i} and H$\alpha$ line profiles at K1 and K2 averaged over an area of 3$^{\prime\prime}$$\times$3$^{\prime\prime}$. The \emph{vertical dotted line} in each panel denotes the line center of Fe \textsc{i} or H$\alpha$.}\label{fig:3}
\end{figure}

Figures \ref{fig:3}a and b show the temporal evolution of the relative enhancements of continuum and spectral line emissions from HMI, WST, and CHASE, which are integrated over an area of 3$^{\prime\prime}$$\times$3$^{\prime\prime}$ for the two white-light kernels K1 and K2. One can see that at K1 (Figure \ref{fig:3}a), all the enhancement curves exhibit an evident rise and then a decay during the flare. The maximum enhancements or increases of HMI, CHASE, and WST continua at $\approx$04:10 UT are 3.9\%, 3.2\%, and 4.7\%, respectively. For comparison, the CHASE Fe \textsc{i} line core and integrated H$\alpha$ line over $\pm$2.3 \AA\ have larger maximum increases of 6.4\% and 38.0\%, respectively. At K2 (Figure \ref{fig:3}b), the maximum increases of HMI, CHASE, and WST continua, as well as CHASE Fe \textsc{i} core and H$\alpha$ line for their main peaks at $\approx$04:10 UT are 4.3\%, 3.0\%, 1.9\%, 2.2\%, and 20.2\%, respectively, most of which are smaller than the corresponding values at K1. Note that the other peaks in the HMI and CHASE continua, say, before 04:08 UT and after 04:15 UT, might be unrelated to the flare itself, as no obvious response is found in the HXR emissions. 

Figures \ref{fig:3}c\,--\,f exhibit the temporal evolution of CHASE Fe \textsc{i} and H$\alpha$ line profiles averaged over an area of 3$^{\prime\prime}\times$3$^{\prime\prime}$ at K1 and K2. One can see that at both kernels, the Fe \textsc{i} line remains an absorption profile with a symmetric Gaussian shape during the flare (Figures \ref{fig:3}c\,--\,d). Its line core as well as the nearby continuum have an increase followed by a decrease in the intensity over time. The H$\alpha$ line varies from an absorption to an emission and then back to an absorption profile (Figures \ref{fig:3}e\,--\,f). Its emission profiles display a central reversal with double peaks in line wings, which are not symmetric. K1 shows a red asymmetry but K2 has a blue asymmetry, namely the conjugate footpoints have an opposite asymmetry in H$\alpha$. 

\begin{figure} 
\centerline{\includegraphics[width=1 \textwidth]{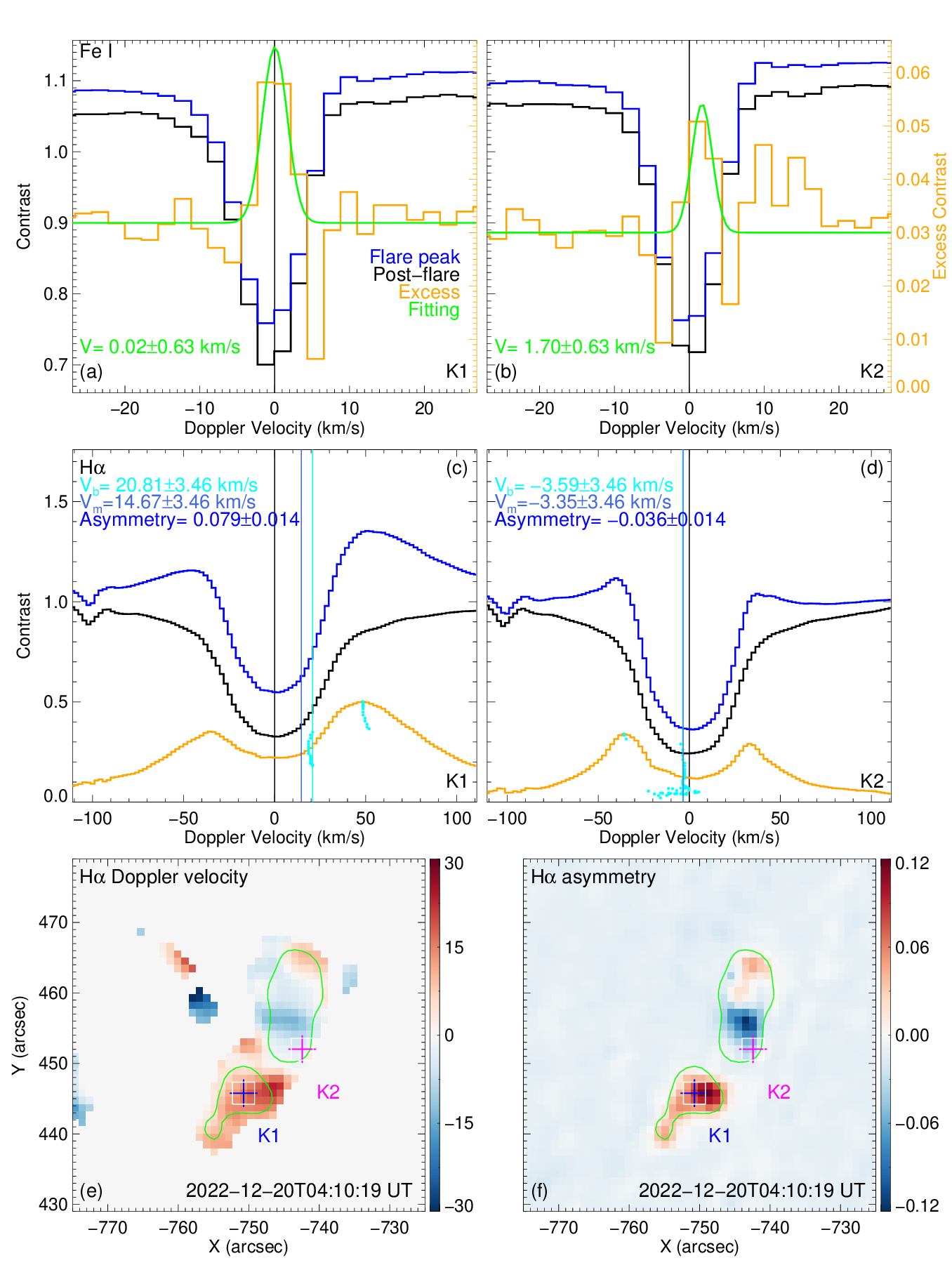}}
\caption{Normalized line profiles of Fe \textsc{i} (\textbf{a} and \textbf{b}) and H$\alpha$ (\textbf{c} and \textbf{d}) from the white-light kernels K1 and K2 averaged over an area of 3$^{\prime\prime}$$\times$3$^{\prime\prime}$ at the flare peak time of 04:10:19 UT. The \emph{blue}, \emph{black}, and \emph{orange curves} in each panel indicate the flaring, background, and excess profiles, respectively. The \emph{gray vertical line} in each panel denotes the line center of Fe \textsc{i} or H$\alpha$. The \emph{green curve} in Panels \textbf{a} and \textbf{b} represents a Gaussian fit to the excess Fe \textsc{i} profile. The \emph{royal blue} and \emph{cyan vertical lines} in Panels \textbf{c} and \textbf{d} mark the Doppler velocities derived from the moment and bisector methods, respectively, the latter of which is a medium of the velocities obtained at different intensity levels. (\textbf{e}) and (\textbf{f}) Maps of H$\alpha$ Doppler velocity and asymmetry at the flare peak time. The Doppler velocity is calculated by the moment method. The \emph{two pluses} in each panel represent the two white-light kernels K1 and K2. The \emph{green contours} denote the H$\alpha$ flare ribbons whose excess intensities are larger than 500 DNs.}\label{fig:4}
\end{figure}

We further select the Fe \textsc{i} and H$\alpha$ line profiles with maximum intensities at 04:10:19 UT to derive the Doppler velocity as shown in Figures \ref{fig:4}a\,--\,d. Note that the excess profiles are adopted to do a Gaussian fit for Fe \textsc{i} and to make a moment plus a bisector analysis for H$\alpha$. One can see that the Fe \textsc{i} line at K1 exhibits a trivial redshift velocity which is actually within the uncertainty (Figure \ref{fig:4}a). By contrast, the excess Fe \textsc{i} emission at K2 has an evident redshift with a velocity of 1.70 km s$^{-1}$ (Figure \ref{fig:4}b), indicative of a plasma downflow at the lower atmospheric layer. The excess H$\alpha$ profiles with a red or blue asymmetry at K1 and K2 show a much larger redshift or blueshift velocity (Figures \ref{fig:4}c\,--\,d). At K1, the measured Doppler velocity is 14.67/20.81 km s$^{-1}$ (i.e. redshift) from a moment/bisector method. Meanwhile at K2, the velocity is $-$3.35/$-$3.59 km s$^{-1}$ (blueshift) via moment/bisector. Note that the velocity from the bisector is a median of the ones obtained at different intensity levels. 

Moreover, we derive the H$\alpha$ Doppler velocity map via moment for the time at 04:10:19 UT as shown in Figure \ref{fig:4}e. To get a reliable result, we artificially set the velocity to 0 when the peak intensity of the excess profile in a pixel is less than 100 DNs. It is seen that evident Doppler velocities mainly come from the two brightened flare ribbons as marked by green contours (same as the white contours in Figure \ref{fig:1}h), whose excess intensities are larger than 500 DNs. In particular, H$\alpha$ can show redshift and blueshift velocities at the conjugate footpoints, which has been revealed from the profiles at K1 and K2 as described above. We also plot the asymmetry map in Figure \ref{fig:4}f. One can see that only at the flare ribbons, the H$\alpha$ profiles show a notable red or blue asymmetry with an obvious line-wing enhancement. Note that the asymmetry distribution looks quite similar to that of Doppler velocity at the two ribbons, which can also be revealed by the scatter plot in Figure \ref{fig:6}i. However, one should caution that the red (blue) asymmetry of H$\alpha$ profiles, or the measured redshift (blueshift) velocity is not necessarily caused by a plasma downflow (upflow), as already revealed in previous studies \citep[e.g.][]{Gan1993, Heinzel1994, Ding1996, Kuridze2015}. 
It is also interesting to see that the asymmetry is reversed for the conjugate footpoints at K1 and K2. As given in Figures \ref{fig:4}c and d, the asymmetries of H$\alpha$ profiles at K1 and K2 are 0.079 and $-$0.036, respectively. 

\subsection{Photospheric Magnetic Field Changes at the White-light Kernels}
  \label{S-deltab} 
  
\begin{figure} 
\centerline{\includegraphics[width=1 \textwidth]{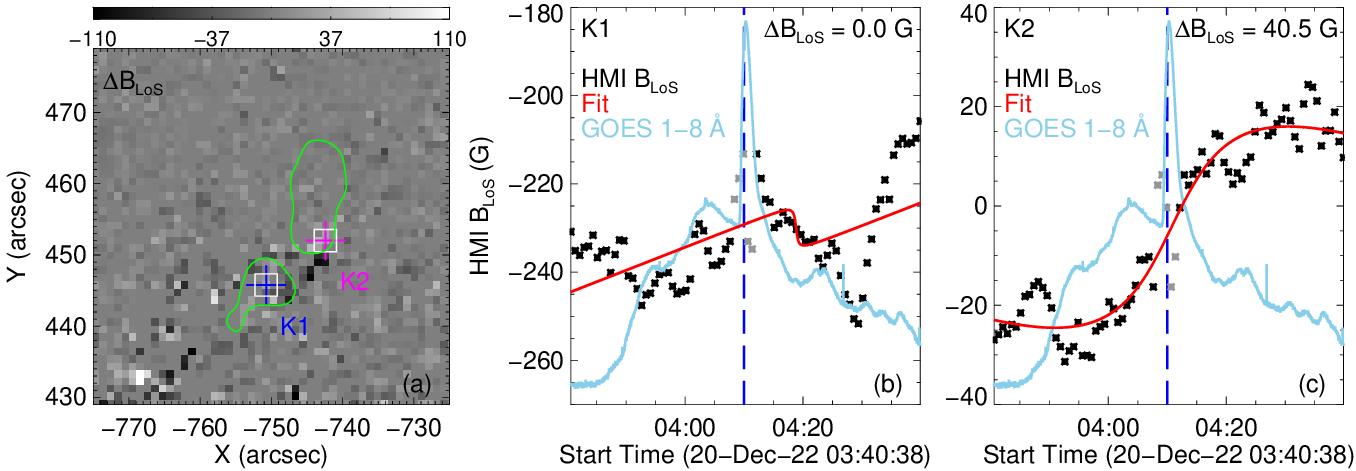}}
\caption{(\textbf{a}) Photospheric $\Delta B_{\rm LoS}$ map for the flare active region. The \emph{two pluses} indicate the two white-light kernels K1 and K2 and the \emph{green contours} denote the H$\alpha$ ribbons whose excess intensities are larger than 500 DNs. (\textbf{b}) and (\textbf{c}) Temporal evolution of HMI $B_{\rm LoS}$ (\emph{black dots}) at K1 and K2 averaged over an area of 3$^{\prime\prime}$$\times$3$^{\prime\prime}$, together with the GOES SXR 1\,--\,8 \AA\ light curve (\emph{sky blue curve}). In each panel, the \emph{red line} denotes a stepwise function fit to $B_{\rm LoS}$ excluding the five data points (\emph{gray dots}) around the flare peak time (marked by the \emph{blue vertical line}).}\label{fig:5}
\end{figure}

Figure \ref{fig:5}a shows the map of photospheric magnetic field change, i.e., $\Delta B_{\rm LoS}$, for the flare region. We can see some notable changes in $B_{\rm LoS}$ (high $\Delta B_{\rm LoS}$) near/at the white-light kernels K1 and K2. Note that some prominent changes of $B_{\rm LoS}$ also appear in some other regions without white-light brightenings, which is actually similar to the study of \cite{Castellanos_Duran2020}. We further plot the temporal evolution of $B_{\rm LoS}$ at K1 and K2 as well as the fit using a stepwise function in Figures \ref{fig:5}b and c. It is seen that there seems to be no obvious change in $B_{\rm LoS}$, i.e. with a poor fit, at K1, while at K2, there exists an evident change in $B_{\rm LoS}$ with a value of 40.5 Gauss during the flare. In fact, the relationship between the white-light emission and $B_{\rm LoS}$ in this small WLF is somewhat similar to the results in \cite{Castellanos_Duran2020}.  

\subsection{Statistical Relationships among Different Parameters from Flare Ribbons} 
  \label{S-correlation}
  
\begin{figure} 
\centerline{\includegraphics[width=1 \textwidth]{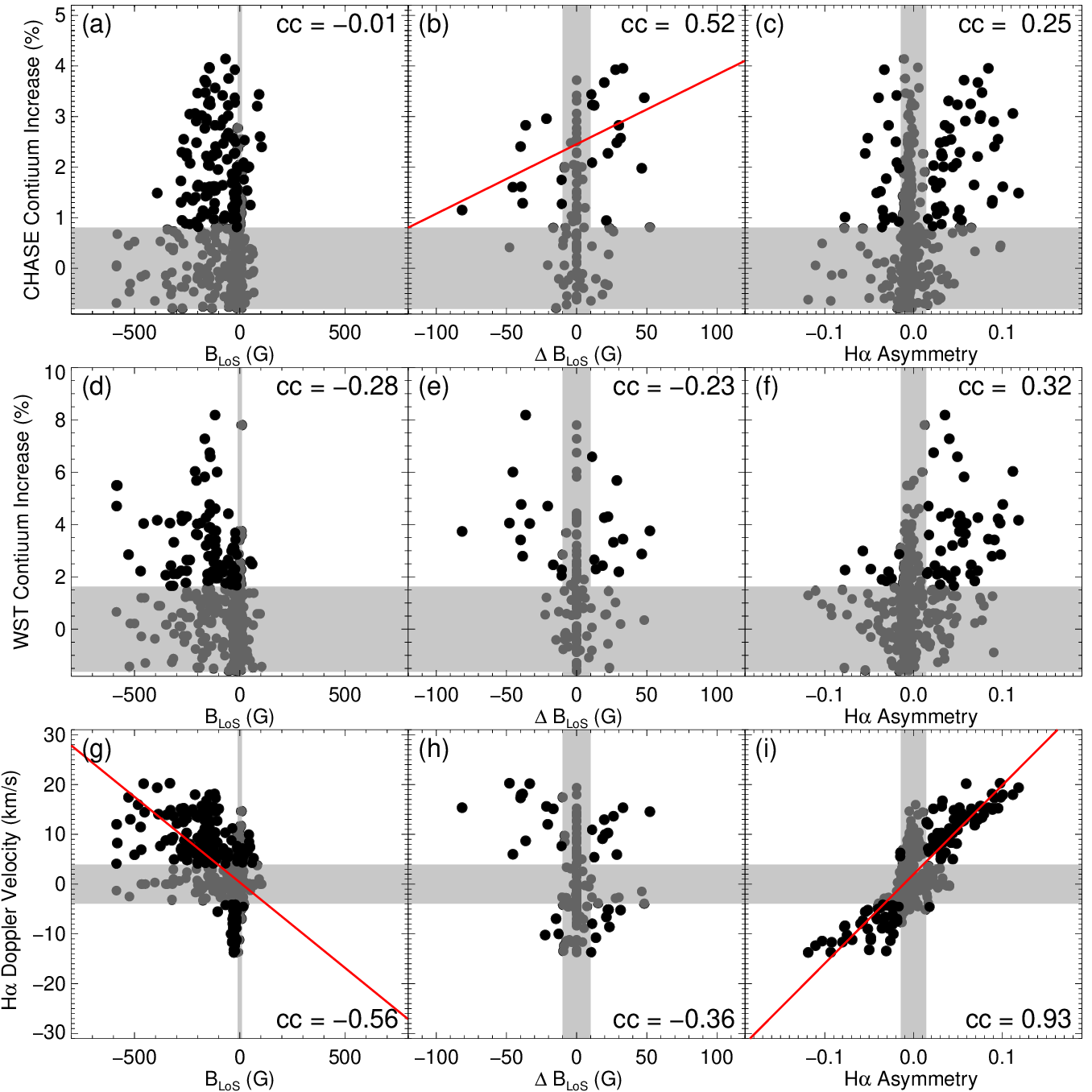}}
\caption{(\textbf{a\,--\,c}) Scatter plots of CHASE continuum increase with HMI $B_{\rm LoS}$, $\Delta B_{\rm LoS}$, and H$\alpha$ asymmetry from flare ribbons. (\textbf{d\,--\,f}) scatter plots of WST 3600 \AA\ increase with the above mentioned three parameters. (\textbf{g\,--\,i}) Scatter plots of H$\alpha$ Doppler velocity with the three parameters. The \emph{gray area} in each panel denotes the uncertainty of each parameter. We do a linear fit (\emph{red line}) if two parameters among them have a good correlation (cc$\ge0.5$).}\label{fig:6}
\end{figure}

We check the statistical relationships among different parameters within the two flare ribbons (marked by the contours in Figures \ref{fig:1}h, \ref{fig:4}e and f, and \ref{fig:5}a). To increase the reliability, here we collect all the available time frames in observations, namely four frames for the CHASE continuum around its maximum, WST continuum, HMI $B_{\rm LoS}$, and CHASE H$\alpha$ asymmetry but only one frame for HMI $\Delta B_{\rm LoS}$. Figures \ref{fig:6}a\,--\,c show the scatter plots of CHASE continuum increase versus photospheric $B_{\rm LoS}$, $\Delta B_{\rm LoS}$, and H$\alpha$ asymmetry. Note that the data points in the shaped area, i.e. within uncertainties, are excluded to further calculate the linear Pearson correlation coefficient (cc). It is seen that the CHASE continuum increase only has some correlation with $\Delta B_{\rm LoS}$ (cc=0.52). In contrast, it has no obvious relationship with $B_{\rm LoS}$ (cc=$-$0.01) or H$\alpha$ asymmetry (cc=0.25). Figures \ref{fig:6}d\,--\,f demonstrate the scatter plots of WST continuum increase versus $B_{\rm LoS}$, $\Delta B_{\rm LoS}$, and H$\alpha$ asymmetry. We find that the continuum increase at WST 3600\,\AA\ has no linear relationships with $B_{\rm LoS}$ (cc=$-$0.28), $\Delta B_{\rm LoS}$ ($-$0.23), or H$\alpha$ asymmetry (cc=0.32). Figures \ref{fig:6}g\,--\,i exhibit the scatter plots of H$\alpha$ Doppler velocity versus $B_{\rm LoS}$, $\Delta B_{\rm LoS}$, and H$\alpha$ asymmetry. We can see that H$\alpha$ Doppler velocity has good relationships with $B_{\rm LoS}$ (cc=$-$0.56) as well as H$\alpha$ asymmetry (cc=0.93). By contrast, it has no good correlation with $\Delta B_{\rm LoS}$ (cc=$-$0.36). Note that the negative correlation between H$\alpha$ velocity and $B_{\rm LoS}$ is owing to negative magnetic fields at most ribbon locations. The implications or explanations of these relationships among different parameters will be discussed in the following Section.

\section{Summary and Discussions}
     \label{S-Summary} 

In this paper, we perform a comprehensive study of a small C2.3 WLF observed by CHASE and ASO-S, combining with the multi-band observations from SDO and SolO. Our main results are summarized as follows.

\begin{enumerate}
\item This small flare shows relative enhancements or increases of 4.3\%, 3.2\%, 4.7\%, and 6.4\% in the HMI, CHASE, and WST continua and CHASE Fe \textsc{i} line, respectively. The white-light brightening kernels are mainly located near a small pore as well as within the H$\alpha$ flare ribbons.
\item At the white-light kernels, the CHASE Fe \textsc{i} line remains an absorption profile with a symmetric Gaussian shape during the flare. It can show a redshift velocity up to 1.7 km s$^{-1}$. By contrast, the H$\alpha$ line changes from an absorption to an emission profile, the latter of which has a central reversal with a line-wing enhancement. In addition, the H$\alpha$ emission profile displays a red or blue asymmetry corresponding to a velocity of several to tens of km s$^{-1}$. 
\item Due to a low cadence of HMI, CHASE, and WST observations, the peak times of white-light emissions and their temporal relationship with the HXR emission are inconclusive in this flare. However, the white-light kernels are found to be co-spatial with the nonthermal HXR sources at 20\,--\,35 keV. This suggests that the white-light emissions are related to a nonthermal electron-beam heating, whose radiative backwarming effect could not be excluded.
\item One of the identified white-light kernels is accompanied by a change of photospheric magnetic field [$\Delta B_{\rm LoS}$], while the other is not. Only the CHASE continuum increase at the flare ribbons has a good relationship with $\Delta B_{\rm LoS}$. These confirm that the continuum emission can be linked to $\Delta B_{\rm LoS}$ but in an unknown way \citep{Castellanos_Duran2020}.
\item The H$\alpha$ Doppler velocity at the flare ribbons has good correlations with $B_{\rm LoS}$ as well as H$\alpha$ asymmetry in this small WLF. 
\end{enumerate}

From a statistical perspective (Section \ref{S-correlation}), we find a good correlation between the CHASE continuum increase and photospheric $\Delta B_{\rm LoS}$ at flare ribbons, while the case is not between the WST continuum increase at 3600 \AA\ and $\Delta B_{\rm LoS}$. This may be explained by the different formation heights of these two continua. The CHASE continuum near the Fe \textsc{i} line at 6569.2 \AA\ can be formed lower (say in either the photosphere or chromosphere: e.g. \citealt{Kleint2016}; \citealt{kerr2017}) than the WST 3600 \AA\ continuum (i.e. Balmer continuum), the latter of which mainly originates from the chromosphere during a flare \citep[e.g.][]{Avrett1986}. As regards the H$\alpha$ line, it is formed from the photosphere to chromosphere from its wing to core. In particular, this line is sensitive to nonthermal heating (e.g. \citealt{Henoux1993}; \citealt{Rubio_da_Costa2016, Capparelli2017}). Therefore, its Doppler velocity could reflect chromospheric condensation and probably also evaporation \citep[e.g.][]{Song2023} caused by nonthermal electron beams. The H$\alpha$ velocity shows some correlation with the photospheric $B_{\rm LoS}$. This implies that the chromospheric condensation/evaporation can be influenced by the magnetic field in the lower atmosphere particularly in WLFs, as in a WLF, the position of chromospheric condensation/evaporation could be relatively lower in height due to a lower energy deposition height compared with non-WLFs. 

This small C2.3 WLF has some properties similar to those of large or major WLFs. (1) The CHASE Fe \textsc{i} and H$\alpha$ lines at the white-light kernels in this small WLF display similar profiles, i.e. symmetric Gaussian profiles still in absorption in Fe \textsc{i} and asymmetric emission profiles with a central reversal in H$\alpha$, to those in an X1.0 WLF as reported by \cite{Song2023}, although their relative increases in the intensity are much lower compared with the X1.0 WLF. (2) The white-light emissions in this small WLF are related to a nonthermal electron-beam heating, which is the same as the case in many large WLFs \citep[e.g.][]{Kuhar2016, Lee2017, Song2023}. (3) Some white-light brightening points in this small WLF are accompanied by a photospheric $\Delta B_{\rm LoS}$ while some are not. This is consistent with the results in X- and M-class flares \citep{Castellanos_Duran2020}. All these may hint that the small WLFs with C and even B class may be the same as the large WLFs with M and X class in nature. 

More small WLFs are worthy of study in the future, especially with comprehensive observations. Firstly, combining with HMI continuum images at Fe \textsc{i} 6173 \AA\ (in the Paschen continuum) and WST images at 3600 \AA\ (in the Balmer continuum), we can investigate the types of WLFs, i.e. type I or II with or without a Balmer jump \citep[e.g.][]{Fang1995}, as well as the origins of these continuum emissions in small WLFs. In particular, WST provides 3600\,\AA\ images with a very high cadence of one second in a burst mode, which can capture some rapid variations of white-light brightenings during a WLF. Secondly, using the spectroscopic observations of Fe \textsc{i} and H$\alpha$ lines from CHASE, we can obtain the dynamics from the photosphere to chromosphere induced by the energy deposition of a WLF. The Doppler velocity and also line width from conjugate footpoints deserve a further study, which could provide some insights into the heating or energy deposition of a WLF. Thirdly, HXR spectral and imaging observations from HXI and/or STIX are necessary to determine the heating mechanisms of a WLF. Some quantitative relationships between the HXR flux (or even low-energy cutoff or spectral index of the spectrum) and continuum increase further help to establish a detailed relation between the electron-beam heating and white-light emissions in observations. Finally, magnetic field observations from HMI and also the \textit{Full-disk vector MagnetoGraph} (FMG: \citealp{Su2019_raa}) on ASO-S are important to understand how the magnetic field influences the white-light emissions, which has been unknown even in large WLFs. Overall, on the one hand, a statistical study for small WLFs is worthwhile to be performed using high-resolution observations. On the other hand, with the aid of sophisticated radiative hydrodynamic simulations, some case studies are required to understand the energy transportation and deposition as well as emission mechanisms for both large and small WLFs.
     
\begin{acks}
We are very grateful to the anonymous reviewer for his/her valuable suggestions on this manuscript. The ASO-S mission is supported by the Strategic Priority Research Program on Space Science, Chinese Academy of Sciences. The CHASE mission is supported by the China National Space Administration. SDO is a mission of NASA’s Living With a Star Program. Solar Orbiter is a space mission of international collaboration between ESA and NASA, operated by ESA. The STIX instrument is an international collaboration between Switzerland, Poland, France, Czech Republic, Germany, Austria, Ireland, and Italy. We thank Shihao Rao, Xiaoshuai Zhu, Bing Wang, Zhentong Li, and Ye Qiu for their help in data processing.
\end{acks}

\begin{authorcontribution}
Q. Li carried out the data analysis and wrote the first manuscript. Y. Li conceived the idea of the work and revised the manuscript. Y. Su provided the HXI data and imaging. D.C. Song provided suggestions on data analysis. W.Q. Gan is PI of ASO-S. H. Li and L. Feng are PI and Co-PI of LST, respectively. Y. Huang, Y.P. Li, J.W. Li, J. Zhao, L. Lu, B.L. Ying, J.C. Xue, P. Zhang, J. Tian,  X.F. Liu, G. Li, J.C. Jing, S.T. Li, G.L. Shi, Z.Y. Tian, W. Chen, Y.N. Su, Q.M. Zhang, D. Li, Y.Y. Ge, J.H. Shan, Y. Zhou, and S.J. Lei contributed to the pipeline and release of ASO-S data. All authors reviewed the manuscript.
\end{authorcontribution}

\begin{fundinginformation}
The authors are supported by the National Natural Science Foundation of China (NSFC) under grants 12233012 and 12273115, the Strategic Priority Research Program of the Chinese Academy of Sciences under grant XDB0560000, the National Key R\&D Program of China under grant 2022YFF0503004, and NSFC under grants 11921003 and 12203102.
\end{fundinginformation}

\begin{dataavailability}
The ASO-S data used in this work are not publicly available, which were observed during the commissioning phase (from 9 October 2022 to 31 March 2023). However, they are available from the corresponding author on reasonable request. The ASO-S data since 1 April 2023 are publicly available and can be accessed from the official website of ASO-S at \urlurl{aso-s.pmo.ac.cn/sodc/dataArchive.jsp}. SDO data are publicly accessible at \urlurl{jsoc.stanford.edu}. STIX data are publicly available at \urlurl{datacenter.stix.i4ds.net}.
\end{dataavailability}

\begin{ethics}
\begin{conflict}
The authors declare no conflicts of interest
\end{conflict}
\end{ethics}

\bibliographystyle{spr-mp-sola}
\bibliography{reference}  

\end{article} 

\end{document}